\documentclass[10pt]{article}

\begin{document}

\title{Self-similar cosmologies in $5D$: Our universe  as a topological separation from an empty $5D$ Minkowski space. }
\author{J. Ponce de Leon\thanks{E-Mail:
jpdel@ltp.upr.clu.edu, jpdel1@hotmail.com}  \\Laboratory of Theoretical Physics, 
Department of Physics\\ 
University of Puerto Rico, P.O. Box 23343,  
San Juan,\\ PR 00931, USA}
\date{February, 2008}

\maketitle
\begin{abstract}
In this paper we find the most general self-similar, homogeneous and isotropic, Ricci flat cosmologies in $5D$.  
These cosmologies show a number of interesting features: (i) the field equations allow a complete integration in terms of  one arbitrary function of the similarity variable,  and a free parameter; 
(ii) the three-dimensional spatial surfaces are flat;   (iii) the extra dimension is spacelike; (iv) the general solution  is Riemann-flat in $5D$ but curved in $4D$, which means that     
an observer confined to $4D$ spacetime can relate this curvature to the presence of matter,  as determined by the Einstein equations in $4D$.
We show that these cosmologies can be interpreted, or used,  as $5D$ Riemann-flat embeddings for spatially-flat FRW cosmologies in $4D$. In this interpretation our universe arises as a topological separation from an empty $5D$ Minkowski space, as envisioned by Zeldovich.

\end{abstract}

\medskip

PACS: 04.50.+h; 04.20.Cv

{\em Keywords:} Space-Time-Matter theory; General Relativity; Exact solutions; Cosmological models; Self-similar symmetry.

\newpage
\section{Introduction}

The question of whether there was a Beginning of the Universe,  is a truly challenging puzzle for physics and for philosophy \cite{Turok}. In classical four-dimensional  general relativity, the singularity theorems imply that  the big bang was a unique birth event for a universe filled with   matter satisfying the weak, dominant and strong  energy conditions \cite{Hawking}-\cite{Wald}. 
The big bang model postulates that our observable universe originated in a singularity sometime between $10$ and $20$ billion years ago. Before the bang nothing existed, not space, time, matter, or energy.

Astronomical observations support the notion that our universe began at some point in the finite past \cite{Bernandis}-\cite{Spergel2}. The question is {\it how} did it begin. 
Twenty seven years ago, in a paper concerning the idea of a spontaneous birth of our universe, Ya. B. Zeldovich observed that ``there is a certain arbitrariness and fuzziness in the very concept of spontaneous birth" \cite{Zeldovich}. He inquired whether spontaneous birth emerges (i) ``out of nothing",  or (ii) in a space of more dimensions,  or (iii) as a topological separation from an initially given empty Minkowski space.

Today, in the literature we find a huge number of papers examining theories for the birth of our universe along the lines (i) and (ii) envisioned by Zeldovich\footnote{It should be mentioned that in addition to  the theories envisioned by Zeldovich, there are many other ones, e.g., evolution of our universe from a $4D$ Minkowski spacetime \cite{Bonnor}, \cite{WessonB}. More recently, we find black hole generation of universes; universes spontaneously creating other universes; creation of universes by intelligent life. For a wonderful popular review, see for example \cite{Book}}. In fact, there are so many interesting papers and books about these topics that for us here is impossible, and far beyond the scope of this work,   to give a thorough  list of references. Therefore we restrict ourselves to mention just few representative works where one can find a detailed bibliography:    "tunneling from nothing" \cite{Vilenkin1}-\cite{Linde}; spontaneous creation of the braneworld \cite{Gorsky1}-\cite{Firouzjahi}; induced matter in Kaluza-Klein gravity and STM (Space-Time-Matter) theories  \cite{JPdeL1}-\cite{Sanjeev}.

Concerning the third alternative mentioned by Zeldovich,   the hypothesis that singularities in $4D$ are induced by the separation of spacetime from the other dimensions has been examined by the present author \cite{JPdeLgr-qc/0106020}; Seahra and Wesson have thoroughly investigated the structure of the big bang from a five-dimensional embedding \cite{JPdeL1} for the standard spatially-flat $4D$ FRW models \cite{SanjeevWesson}. Besides  these studies, there are only few more works related to this alternative  \cite{Lyndell-Bell}-\cite{Ringermacher}. 
The aim of this paper is to present a general class of spatially homogeneous and isotropic solutions of the Einstein field equations in $5D$ that is  relevant to this  alternative. 

To be more precise, in many Kaluza-Klein and braneworld theories the higher dimensional space is assumed to be either Ricci-flat   or anti-de Sitter. In this work we develop  a family of metrics  that are Ricci-flat and  Riemann-flat in $5D$, but whose four-dimensional subspaces are curved. 
Therefore, they  are all equivalent
to an empty Minkowski space in $5D$. However, since the Riemann tensor of four-dimensional subspaces is non-vanishing, for an observer confined to $4D$ the spacetime is not empty but contains  matter as determined by the Einstein equations in $4D$. 

The solutions arise from the observation that cosmological models are self-similar \cite{JPdeL2}. In order to illustrate this in $5D$, let us consider the  metric\footnote{Conventions: Throughout the paper we use geometric units where $c = G = 1$; $ t = x^0,\; r = x^1,\; \theta = x^2$ and $\phi = x^ 3$ are the usual coordinates for a spacetime with spherically symmetric spatial sections; $y = x^4$ represents the coordinate along the extra dimension; the signature of the $5D$ metric is $(+, -, -, -, \epsilon)$ where $\epsilon$ can be either $- 1$ or $+ 1$ depending on whether the extra dimension is spacelike or timelike. The range of tensor indices is $A, B...= 0-4$ and $\mu, \nu, ... = 0-3$.} \cite{JPdeL1}
\begin{equation}
\label{standard cosmological solution in 5D}
dS^2 = Cy^2 dt^2 - D t^{2/\alpha} y^{2/(1 - \alpha)}\left[dr^2 + r^2\left(d\theta^2 + \sin^2\theta d\phi^2\right)\right] - \frac{C \alpha^2 t^2}{(1 - \alpha)^2}dy^2,
\end{equation}
where $\alpha$ is a free parameter and  $C$ as well as  $D$ are constants with the appropriate units. The properties of this metric have widely been discussed in the literature, e.g.,  \cite{Wesson book}, \cite{SanjeevWesson}, \cite{OverduinWesson}. It is Riemann-flat and apparently empty in $5D$, but reduces to the well-known $4D$ FRW models with flat $3D$ sections on   $y =$ constant hypersurfaces (henceforth denoted by $\Sigma_{y}$). It is easy to verify that (\ref{standard cosmological solution in 5D})  admits a homothetic Killing vector, viz., 
\begin{equation}
\label{homothetic Killing vector 1}
{\cal{L}}_{\xi} g_{AB} = 2 g_{AB}, \;\;\;\mbox{with}\;\;\; \xi^A = \left(\frac{\alpha t}{(2\alpha - 1)},\; {r},\; 0,\; 0,\; \frac{(\alpha - 1)y }{(2\alpha - 1)}\right).
\end{equation}
 With the transformation of coordinates
\begin{equation}
t \rightarrow {\bar{t}}^{\;{\alpha}/{(2\alpha - 1)}},\;\;\;r \rightarrow \bar{r}, \;\;\;y \rightarrow {\bar{y}}^{\;(\alpha - 1)/(2\alpha - 1)},
\end{equation}
we have 
\begin{equation}
\label{5D homothetic vector in self-similar coordinates}
\xi^{A} \rightarrow {\bar{\xi}}^A = \left(\bar{t},\; \bar{r},\; 0,\; 0, \; \bar{y}\right)
\end{equation}
and the metric (\ref{standard cosmological solution in 5D}) becomes
\begin{equation}
\label{standard cosmological solution in 5D in self-similar form}
dS^2 = \bar{A}\; \xi^{2(1 - \alpha)/(2\alpha - 1)}d{\bar{t}}^2 - \bar{B}\; \xi^{2/(2\alpha - 1)}\left[dr^2 + r^2\left(d\theta^2 + \sin^2\theta d\phi^2\right)\right] -  \bar{A}\; \xi^{2\alpha/(2\alpha - 1)} \; d{\bar{y}}^2,
\end{equation}
where $\bar{A}$ and $\bar{B}$ are dimensionless constants and 
\begin{equation}
\label{dimensionless variable for our old solution}
\xi = \frac{\bar{t}}{\bar{y}}.
\end{equation}
The above clearly illustrates the self-similar nature of (\ref{standard cosmological solution in 5D}). In this work we extend our previous studies \cite{JPdeL1} and \cite{JPdeL2} to obtain a general class of self-similar cosmologies  that are Ricci-flat and Riemann-flat in $5D$, but curved in $4D$. Consequently, they can be used, or interpreted, as $5D$ Riemann-flat  embeddings for our $4D$ universe, which is consistent  with Zeldovich's notion that  our universe could have emerged as a topological separation from an initially given empty Minkowski space in more than four dimensions.

The paper is organized as follows. In Section $2$, we deduce the  equations with self-similarity in $5D$. In Section $3$, we obtain the general solutions  to these equations; they contain an arbitrary function of the similarity variable, and a free parameter.  In Section $4$, we focus our attention on a class of  solutions that  admits a particularly simple homothetic Killing vector in $5D$, and study their possible application  in $4D$. We use the flexibility of the solution to develop some cosmological models in $4D$. In Section $5$, we present a summary of our results and propose several extensions for this work. Finally, in the Appendix, we discuss the homothetic symmetry on $\Sigma_{y}$.

\section{Field equations}

In this section we present the field equations for self-similar cosmologies in $5D$. 
We start with  the
 line element for a spacetime that has spatial spherical symmetry 
\begin{equation}
\label{most general line element}
dS^2 = e^{\nu(r, t, y)}dt^2 - e^{\lambda(r, t, y)}dr^2 - R^2(r, t, y)\left[d\theta^2 + \sin^2\theta d\phi^2\right] + \epsilon \Phi^2(r, t, y)dy^2.
\end{equation}
For cosmological models we assume spatial homogeneity, which means that the metric is invariant under spatial translations.  However, we do not make any assumption regarding the curvature of $3D$-space. Now, as illustrated above (\ref{standard cosmological solution in 5D in self-similar form}), by a suitable transformation of coordinates in a self-similar model  all the dimensionless quantities can be put in a form where they are functions only of a single variable (say $\xi$) \cite{Sedov}-\cite{JPdeL5}. Thus, in the case under consideration, in ``self-similar" coordinates $\bar{t}, \bar{r}$, and  $\bar{y}$, the line element (\ref{most general line element}) can be written as 
\begin{equation}
\label{most general self-similar line element}
dS^2 = e^{\nu(\xi)}d{\bar{t}}^2 - e^{\lambda(\xi)}d{\bar{r}}^2 - {\bar{r}}^2e^{\mu(\xi)}\left[d\theta^2 + \sin^2\theta d\phi^2\right] + \epsilon \Phi^2(\xi)d{\bar{y}}^2.
\end{equation}
On the other hand, we have {\it no} reason to set $ \xi = \bar{t}/\bar{y}$ as in (\ref{dimensionless variable for our old solution}). In principle $\xi$ can be any function of $\bar{t}$ and $\bar{y}$, namely
\begin{equation}
\label{The most general self-similar variable}
\xi = \xi(\bar{t}, \bar{y}).
\end{equation}
 With this choice the line element (\ref{most general self-similar line element}) is self-similar but not necessarily admits a homothetic Killing vector (More comments about this at the end of Section $3$). In what follows we are going to suppress the bar over the self-similar coordinates. 

The metric functions in (\ref{most general line element}) and (\ref{most general self-similar line element}) have to satisfy the $5D$ Einstein field equations in apparent vacuum, which in terms of the Ricci tensor are
\begin{equation}
\label{field equations}
R_{AB} = 0.
\end{equation}
At once we note  that ${{R}}_{01} = 0$ requires\footnote{In what follows $f_{\xi}$ denotes derivative of $f$ with respect to $\xi$; dots and primes stand for  derivatives with respect to $t$ and $y$, respectively.}
\begin{equation}
\label{R01}
\lambda_{\xi} = \mu_{\xi}
\end{equation}
 from which we get 
\begin{equation}
e^{\mu(\xi)} = C_{0}e^{\lambda(\xi)},
\end{equation}
where $C_{0}$ is a constant of integration. Now, from ${R}_{1}^{1} = {R}_{2}^{2} = {R}_{3}^{3}$ it follows that 

\begin{equation}
\label{S0}
C_{0} = 1.
\end{equation}
Consequently, the requested self-similarity (\ref{most general self-similar line element}), together with the field equations (\ref{field equations}), demand $R = re^{\lambda/2}$. Therefore, the line element for self-similar cosmological models in $5D$ reduces to
\begin{equation}
\label{most general self-similar line element 1}
dS^2 = e^{\nu(\xi)}dt^2 - e^{\lambda(\xi)}\left[dr^2 + r^2\left(d\theta^2 + \sin^2\theta d\phi^2\right)\right] + \epsilon \Phi^2(\xi)dy^2.
\end{equation}
This implies that in self-similar  cosmologies the $3D$ spatial  sections ($t$ = constant, $y = $ constant) are flat, in agreement  with astrophysical data \cite{Bernandis}-\cite{Spergel2}. 
For this metric the Ricci tensor in $5D$ has  four non-trivial, independent, components. These are 
\begin{equation}
{{R}}_{00},\;\;\;{{R}}_{04},\;\;\; {{R}}_{11} = \frac{{{R}}_{22}}{r^2} = \frac{{{R}}_{33}}{r^2 \sin^2\theta},\;\;\; {{R}}_{44}.
\end{equation}
A simple analysis of equation $R_{0 4} = 0$, indicates that self-similarity requires that the ratio $[\dot{\xi}'/({\xi}' \dot{\xi})]$ be some function of  $\xi$. Clearly, {\it any} separable function of $t$ and $y$ will do the job. Therefore, without loss of generality  we can set
\begin{equation}
\label{general xi}
\xi = \frac{T({t})}{Y({y})},
\end{equation}
where $T$ and $Y$ are some functions to be determined by the field equations in $5D$. Consequently, we have 
 four differential equations for five unknown, viz.,
\begin{equation}
\nu(\xi),\;\;\; \lambda(\xi),\;\;\;\Phi(\xi),\;\;\;T(t),\;\;\;Y(y).
\end{equation}
We note that $\epsilon$ is also to be defined from the field equations. We will see that  we can solve the field equations (\ref{field equations}) in terms of one arbitrary function and a free parameter. The four equations to be integrated are:
\begin{enumerate}

 \item ${{R}}_{00} = 0$,
\begin{eqnarray}
\label{R00}
\epsilon \Phi^2 {\dot{T}}^2\left(6 \lambda_{\xi\xi} + 3 \lambda_{\xi}^2  - 3\nu_{\xi}\lambda_{\xi} + \frac{4\Phi_{\xi\xi}}{\Phi} - \frac{2\nu_{\xi}\Phi_{\xi}}{\Phi}\right)
+ 2 \epsilon\Phi^2 \ddot{T}Y\left(3\lambda_{\xi} + \frac{2\Phi_{\xi}}{\Phi}\right) +\nonumber \\
  Y'^2 e^{\nu}\xi^2\left(2 \nu_{\xi\xi} + \nu_{\xi}^2 + 3 \nu_{\xi}\lambda_{\xi}  - \frac{2\nu_{\xi}\Phi_{\xi}}{\Phi}\right) + 2 Y'^2  e^{\nu}\left(2 - \frac{Y Y''}{Y'^2}\right)\xi\nu_{\xi} = 0,
\end{eqnarray}

\item ${{R}}_{04} = 0$,

\begin{equation}
\label{R04}
2  \lambda_{\xi\xi} + \lambda_{\xi}^2  +  \frac{2\lambda_{\xi}}{\xi} -  \nu_{\xi}\lambda_{\xi} - \frac{2  \lambda_{\xi}\Phi_{\xi}}{\Phi} = 0.
\end{equation}

\item ${{R}}_{11} = {{{R}}_{22}}/{r^2} = {{{R}}_{33}}/{(r^2 \sin^2\theta}) = 0$,
\begin{eqnarray}
\label{R11}
\epsilon \Phi^2 {\dot{T}}^2\left(2 \lambda_{\xi\xi} + 3 \lambda_{\xi}^2  - \nu_{\xi}\lambda_{\xi}  + \frac{2\lambda_{\xi}\Phi_{\xi}}{\Phi}\right)
+ 2 \epsilon\Phi^2 \ddot{T}Y \lambda_{\xi}+\nonumber \\
   Y'^2 e^{\nu}\xi^2\left(2 \lambda_{\xi\xi} + 3\lambda_{\xi}^2+  \nu_{\xi}\lambda_{\xi}  - \frac{2\lambda_{\xi}\Phi_{\xi}}{\Phi}\right) + 2  Y'^2  e^{\nu}\left(2 - \frac{Y Y''}{Y'^2}\right)\xi \lambda_{\xi} = 0.
\end{eqnarray}
\item Finally, ${{R}}_{44} = 0$, 
\begin{eqnarray}
\label{R44}
 2\epsilon \Phi^2  {\dot{T}}^2\left(\frac{2\Phi_{\xi\xi}}{\Phi}  - \frac{\nu_{\xi}\Phi_{\xi}}{\Phi}  + \frac{3\lambda_{\xi}\Phi_{\xi}}{\Phi}\right) +  4\epsilon \Phi^2 \ddot{T}Y\left(\frac{\Phi_{\xi}}{\Phi}\right) + \nonumber \\
e^{\nu} Y'^2\xi^2\left(2\nu_{\xi\xi} + \nu_{\xi}^2   - \frac{2 \nu_{\xi}\Phi_{\xi}}{\Phi} + 6 \lambda_{\xi\xi} + 3\lambda_{\xi}^2  - \frac{6 \lambda_{\xi}\Phi_{\xi}}{\Phi}\right)
+ 2 Y'^2 e^{\nu}\left(2 - \frac{YY''}{Y'^2}\right)\xi\left(\nu_{\xi} + 3\lambda_{\xi}\right) = 0.
\end{eqnarray}

\end{enumerate}

\section{Integrating the field equations}

Let us first notice that (\ref{R04}) can be easily integrated as 
\begin{equation}
\label{first integral for lambda}
\xi^2 \lambda_{\xi}^2 = C^2 \Phi^2 e^{(\nu - \lambda)},
\end{equation}
where $C$ is a constant of integration. On the other hand, the field equations (\ref{R00}), (\ref{R11}) and (\ref{R44}) have the following structure
\begin{equation}
\label{structure of the equations}
\left(\frac{{\dot{T}}^2}{Y'^2}\right)F(\xi) + \left(\frac{\ddot{T} Y}{Y'^2}\right)G(\xi) + H(\xi) + I(\xi)\left(2 -  \frac{YY''}{Y'^2}\right) = 0,
\end{equation}
where $F$, $G$, $H$ and $I$ symbolize the corresponding functions of $\xi$ in these equations. Therefore, in order to preserve the self-similar symmetry,  we have to require 
\begin{equation}
2 -  \frac{YY''}{Y'^2} = l,
\end{equation}
where $l$ is a separation constant.
Integrating this expression we obtain
\begin{equation}
Y' = \alpha Y^{(2 -l)},
\end{equation}
where $\alpha$ is a constant of integration. Thus, we find 
\begin{equation}
\left(\frac{{\dot{T}}^2}{Y'^2}\right) = \left[\frac{{\dot{T}}^2}{\alpha^2 T^{(4 - 2l)}}\right]\xi ^{(4 - 2l)},\;\;\;\mbox{and}\;\;\;\left(\frac{\ddot{T} Y}{Y'^2}\right) = \left[\frac{\ddot{T}}{\alpha^2 T^{(3 - 2l)}}\right]\xi^{(3 - 2l)}.
\end{equation}
Consistency of (\ref{structure of the equations}) demands the  quantities inside the square brackets to be constants, which requires
\begin{equation}
T \sim \left\{\begin{array}{cc}
            t^{1/(l - 1)}    & \mbox{for $l \neq 1$}, \\
\\
               e^{\beta t}  & \mbox{for $l = 1$},
               \end{array}
      \right.
\end{equation}
where $\beta$ is some constant. 
Consequently, without loss of generality we can set
\begin{equation}
\label{general choice of xi}
\xi = \left\{\begin{array}{cc}
            \left(\frac{t}{y}\right)^{1/(l - 1)}    & \mbox{for $l \neq 1$}, \\
\\
               \left(\frac{e^{\beta t}}{e^{\alpha y}}\right)  & \mbox{for $l = 1$}.
               \end{array}
      \right.
\end{equation}

Let us now calculate  $({{R}}_{00} - {{R}}_{44})$, and use (\ref{R04}) to express the second derivative $\lambda_{\xi\xi}$ in terms of the first derivatives of $\nu$ and $\Phi$. As a result we obtain
\begin{equation}
\label{difference between R00 and R44}
\epsilon \Phi^2\left({T \ddot{T} - {\dot{T}}^2}\right) = \left({Y'^2 - Y Y''}\right)\xi^{2}e^{\nu}.
\end{equation}

\paragraph{Solution for $l \neq 1$:} In this case (\ref{difference between R00 and R44})  yields
\begin{equation}
\label{Phi for k neq 1}
\Phi^2 = (- \epsilon) \xi^{2(l - 1)}e^{\nu}.
\end{equation}
From this expression, it follows that the extra dimension should be spacelike, i.e., $\epsilon = - 1$. 
Now, from (\ref{first integral for lambda}) and (\ref{Phi for k neq 1}) we get
\begin{equation}
\label{enu in terms of lambda}
e^{\nu} = \left(\frac{1}{C}\right)\xi^{(2 - l)}\lambda_{\xi}e^{\lambda/2}.
\end{equation}
In summary, we have found that the line element (\ref{most general self-similar line element 1}) with 
\begin{equation}
\label{self-similar first solution}
e^{\nu} = \left(\frac{1}{C}\right) \xi^{(2 - l)}\lambda_{\xi}e^{\lambda(\xi)/2},\;\;\; \Phi^2 = \xi^{2(l - 1)}e^{\nu(\xi)}, \;\;\;\xi = \left(\frac{t}{y}\right)^{1/(l - 1)}, \;\;\;\mbox{and}\;\;\;\epsilon = - 1,
\end{equation}
is a solution of the $5D$ Ricci-flat equations for any value of the parameter  $l \neq 1$, constant $C$, and arbitrary function $\lambda = \lambda(\xi)$. In addition, it admits a homothetic Killing vector in $5D$, viz., 
\begin{equation}
\label{Lie derivative of the 5D metric}
 {\cal{L}}_{\xi}g_{AB} = 2 g_{AB}, \;\;\;\mbox{with} \;\;\;\xi^{A} = (t, r, 0, 0, y). 
\end{equation}
It should be noted that (\ref{self-similar first solution}) includes the family of $5D$ metrics given by (\ref{standard cosmological solution in 5D}), or (\ref{standard cosmological solution in 5D in self-similar form}) for the particular choice $e^{\lambda(\xi)} = B \xi^{2/(2\alpha - 1)}$ and $l = 2$.

\paragraph{Solution for $l = 1$:} In this case the field equations require $\beta = \alpha$. Therefore, the solution is  

\begin{equation}
\label{self-similar first solution 2}
e^{\nu} = \left(\frac{1}{C}\right) \xi\lambda_{\xi}e^{\lambda(\xi)/2},\;\;\; \Phi^2 = e^{\nu(\xi)}, \;\;\;\xi = \left(\frac{e^{\alpha t}}{e^{\alpha y}}\right), \;\;\;\mbox{and}\;\;\;\epsilon = - 1.
\end{equation}
It is obvious that this solution does not admit a {\it simple} $5D$ homothetic vector as (\ref{homothetic Killing vector 1}) or (\ref{Lie derivative of the 5D metric}) (See bellow). 

\subsection{The Riemann tensor in $5D$}

It can be verified that, for both solutions, all the components of the $5D$ Riemann tensor $R_{ABCD}$ vanish identically. Therefore, they are equivalent to an empty Minkowski space in $5D$ $({\cal{M}}_{5})$. Consequently, there exist some coordinate transformation 
 
\begin{equation}
\label{transformation of coordinates}
\tau = \tau(t, r, y), \;\;\;R = R(t, r, y), \;\;\, \psi = \psi(t, r, y),  
\end{equation}
that brings  (\ref{self-similar first solution}) and (\ref{self-similar first solution 2}) to the line element in ${\cal{M}}_{5}$ 
\begin{equation}
\label{Minkowski space in 5D}
dS^2 = \eta_{AB} dx^A dx^B = d\tau^2 - dR^2 - R^2\left(d\theta^2 + \sin^2\theta d\phi^2\right) - d\psi^2,
\end{equation}
in  Minkowski coordinates
$x^A = (\tau, \; R, \; \theta, \; \phi, \; \psi)$. For the particular solution (\ref{5D homothetic vector in self-similar coordinates}), the explicit coordinate transformation from (\ref{standard cosmological solution in 5D}) to (\ref{Minkowski space in 5D}) is known and has been amply discussed in the literature, see for example \cite{SanjeevWesson} and references therein.

Before going on, and in order to avoid  misunderstandings,   we should comment about the  homothetic nature of the above solutions. Certainly, any  $5D$ vector of the form 
\begin{equation}
\label{Homothetic vector for the Minkowski metric}
\xi^A_{{\cal{M}}} = \left\{[\tau + f_{1}(R, \psi)], \; R, \; 0, \; 0, \; [\psi + f_{2}(\tau, R)]\right\}, 
\end{equation}
where $f_{1}$ and $f_{2}$ are arbitrary functions of their arguments, is a homothetic Killing vector for (\ref{Minkowski space in 5D}), viz., ${\cal{L}}_{\xi_{\cal{M}}} \eta_{AB} = 2\eta_{AB}$.
Thus, the Riemann-flat solutions (\ref{self-similar first solution}) and (\ref{self-similar first solution 2}) admit an infinite number of homothetic killing vectors, which correspond to the nondenumerable infinity of choices of $f_{1}$ and $f_{2}$ in (\ref{Homothetic vector for the Minkowski metric}).

In this context, the difference between (\ref{self-similar first solution}) and (\ref{self-similar first solution 2}) is that, among all the possible choices, in self-similar coordinates $t$, $r$ and $y$ there exists a     very simple homothetic killing vector for the first solution , namely $\xi^{A} = (t, r, 0, 0, y)$, while for the second solution the homothetic vectors (in self-similar coordinates) are much more complicated.

\section{Interpretation in $4D$}
In this section we  discuss possible four-dimensional interpretations of the $5D$ line element (\ref{self-similar first solution}). For simplicity we set 

\begin{equation}
l = 2,
\end{equation} 
in such a way that the similarity variable takes the form $\xi = t/y$, as in (\ref{dimensionless variable for our old solution}). Thus, in what follows we consider the $5D$ metric

\begin{equation}
\label{solution for k = 2}
dS^2 = \left(\frac{1}{C}\right)\lambda_{\xi}e^{\lambda(\xi)/2}\;dt^2 - e^{\lambda(\xi)}\left[dr^2 + r^2\left(d\theta^2 + \sin^2\theta d\phi^2\right)\right] - \left(\frac{1}{C}\right)\xi^2\lambda_{\xi}e^{\lambda(\xi)/2}\;dy^2, 
\end{equation}
with
\begin{equation}
\label{self-similar variable for k = 2}
\xi = \frac{t}{y}.
\end{equation}

The first important question is how  to recover our $4D$ spacetime from $5D$. The most popular approach is to assume that our $4D$ spacetime is a  hypersurface $\Sigma_{y}: y = y_{0}$ = constant, which is orthogonal to the $5D$ unit-vector 
\begin{equation}
n^A = \frac{\delta^A_{4}}{\Phi}
\end{equation}
along the extra dimension, although a dynamical foliation is also possible \cite{JPdeLDynamicalFoliationI}, \cite{JPdeLDynamicalFoliationII}.  The second important question is how  to construct the metric of the physical spacetime from the one induced on $\Sigma_{y}$. There are various approaches in the literature: (i) the canonical metric, which assumes $\Phi = 1$ and factorizes the $4D$ part of the $5D$ metric by an $y^2$ term \cite{Mashhoon2}-\cite{Seahra2}; (ii) the conformal approach, where the $4D$ part of the $5D$ metric is factorized by an $\Phi^N$ term (see for example \cite{ExtSpacetimeForStarsIn5DKK} and references therein); and (iii) the  one where the spacetime metric is identified with  the metric induced on $\Sigma_{y}$. 

An exhaustive treatment of all possibilities is beyond the scope of this work. We present here an introductory analysis where we follow the approach (iii) mentioned above. The induced metric $h_{\alpha\beta}$ on hypersurfaces $\Sigma_{y}$ is just the $4D$ part of the $5D$ metric (\ref{solution for k = 2}), viz.,
\begin{equation}
\label{4D metric}
ds^2 = h_{\mu\nu} dx^{\mu}dx^{\nu} = \left(\frac{1}{C}\right) \lambda_{\xi}e^{\lambda(\xi)/2}dt^2 - e^{\lambda(\xi)}\left[dr^2 + r^2\left(d\theta^2 + \sin^2\phi\right)\right].
\end{equation}
We immediately notice two things:

\medskip

 Firstly, that the spacetime metric $h_{\alpha \beta}$ is {\it not} self-similar along  $\xi^{\mu}_{p} = (t, r, 0, 0)$, which is  the projection of $\xi^{A} = (t, r, 0, 0, y)$ on $\Sigma_{y}$, i.e.,  
\begin{equation}
\label{lost of symmetry in 4D}
{\cal{L}}_{\xi_{p}} h_{\mu\nu} \neq 2h_{\mu\nu}, \;\;\;\mbox{for}\;\;\;{{\xi}}_{p}^{\lambda} = \left({t},\; {r},\; 0,\; 0\right).
\end{equation}
In fact, we show in the Appendix that there is only one family of homothetic solutions on $\Sigma_{y}$. But the homothetic vector in $4D$, say $\zeta^{\mu}$,  is {\it not } parallel to the projected  $\xi^{\mu}_{p}$. 

\medskip

Secondly, although  (\ref{solution for k = 2}) is Riemann-flat, the hypersurfaces $\Sigma_{y}$ are curved.  Indeed, the non-vanishing components of the  Riemann tensor calculated on $\Sigma_{y}$ are 

\begin{eqnarray}
\label{Riemann tensor in 4D}
R_{0101} &=& \frac{R_{0202}}{r^2} = \frac{R_{0303}}{r^2 \sin^2\theta} = \frac{(2\lambda_{\xi \xi} + \lambda_{\xi}^2)e^{\lambda}}{8y^2}     , \nonumber \\
R_{1212} &=& \frac{R_{1313}}{\sin^2\theta} = \frac{R_{2323}}{r^2 \sin^2\theta} = - \frac{C e^{3\lambda/2}\lambda_{\xi}r^2}{4y^2}.
\end{eqnarray}
Now, the requirement $h_{00} \neq 0$ implies $\lambda_{\xi} \neq 0$. Therefore, $R_{\alpha\beta \mu\nu} \neq 0$. What this means is that an observer, who is confined to making physical measurements in our ordinary spacetime, can explain the curvature of $\Sigma_{y}$ as being produced by  (``effective")  matter whose energy-momentum-tensor $T_{\mu\nu}^{(eff)}$
is given by the Einstein equations in $4D$, viz.,
\begin{equation}
\label{definition of induced matter}
G_{\mu\nu} = 8\pi T_{\mu\nu}^{(eff)} \equiv  \epsilon \left[K_{\mu\nu}' + K \left(K_{\mu\nu} - \frac{K}{2}g_{\mu\nu}\right) - 2\left(K_{\mu\rho}K^{\rho}_{\nu} - \frac{1}{4}g_{\mu\nu} K_{\alpha \beta}K^{\alpha \beta}\right)\right],
\end{equation}
where $K_{\mu\nu} = (\partial g_{\mu\nu}/\partial y)/2$. For the case under consideration, a simple calculation yields\footnote{Here $\rho^{(eff)} \equiv h^{00}T_{00}^{(eff)}$, and $p^{(eff)} \equiv - h^{11}T_{11}^{(eff)} = - h^{22}T_{22}^{(eff)} = - h^{33}T_{33}^{(eff)}$.}
\begin{equation}
\label{effective density}
8 \pi \rho^{(eff)} = \frac{3 C}{4 y^2}\;\lambda_{\xi}e^{- \lambda(\xi)/2},
\end{equation}
\begin{equation}
\label{effective pressure}
8 \pi p^{(eff)} = - \frac{C}{2 y^2}\left(\lambda_{\xi} + \frac{\lambda_{\xi\xi}}{\lambda_{\xi}}\right)e^{- \lambda(\xi)/2}.
\end{equation}
We now proceed to illustrate the above discussion with some examples.

\subsection{Solution with $\Phi =  1$: a Riemann-flat embedding for the de Sitter universe} 

Many authors use the five available degrees of coordinate freedom to set $g_{4\mu} = 0$ and  $\Phi = 1$.  This is the so-called ``Gaussian normal coordinates system" based on $\Sigma_{y}$, where $n^{A}$  is taken to  be geodesic in $5D$. In this system we can easily integrate    
(\ref{self-similar first solution}) to obtain (with $k = 2$)

\begin{equation}
\label{Phi = 1}
dS^2 = \frac{1}{\xi^2}\;dt^2 - \left(B + \frac{C}{\xi}\right)^2\left[dr^2 + r^2 d\Omega^2\right] - dy^2
\end{equation}
where $B$ and $C$ are dimensionless constants. With the  transformation of coordinates\footnote{Setting $B = 0$, and making a  coordinate transformation  $dt/t = - \sqrt{\Lambda(\bar{t})/3}\;d{\bar{t}}$, where the function $\Lambda(\bar{t})$ has units of (length)$^{-2}$,  (\ref{Phi = 1}) can be written as   
 $dS^2 = y^2 \frac{\Lambda(\bar{t})}{3} d{\bar{t}}^2 - C^2y^2e^{2\int{\sqrt{\Lambda(\bar{t})/3}d\bar{t}}}[dr^2 + r^2 d\Omega^2] - dy^2$, which is identical to the $5D$ line element discussed by Bellini \cite{Bellini}.  }
\begin{equation}
\frac{dt}{t} = - \frac{d\tilde{t}}{L},\;\;\;t \rightarrow e^{- \tilde{t}/L},
\end{equation}
where $L$ is some constant length, (\ref{Phi = 1}) becomes
\begin{equation}
\label{Phi = 1 in new coordinates}
dS^2 = \frac{y^2}{L^2} d{\tilde{t}}^2 - \left(B + \tilde{C} y e^{\tilde{t}/L}\right)^2\left[dr^2 + r^2d\Omega^2\right] - dy^2,
\end{equation}
where the new constant $\tilde{C}$ has dimensions of $(length)^{- 1}$. 
On $\Sigma_{y}$ the effective matter is given by
\begin{equation}
8 \pi \rho^{(eff)} = \frac{3 {\tilde{C}}^2 e ^{2 \tilde{t}/L}}{(B + \tilde{C}y e^{\tilde{t}/L})^2}, \;\;\;8 \pi p^{(eff)} = - \frac{\tilde{C} e^{\tilde{t}/L}(2 B + 3\tilde{C} y e^{\tilde{t}/L})}{y (B + \tilde{C}y e^{\tilde{t}/L})^2}.
\end{equation}
We note that 
\begin{equation}
8\pi \rho^{(eff)} \rightarrow - 8 \pi p^{(eff)} = \frac{3}{y^2}\;\;\;\mbox{as}\;\;\;\tilde{t} \rightarrow \infty.
\end{equation}
Therefore (\ref{Phi = 1}), or (\ref{Phi = 1 in new coordinates}), for $t \approx 0$ is a Riemann-flat embedding for the de Sitter universe with cosmological ``constant" $\Lambda = 3/y^2$.

\subsection{Solution with $e^{\nu} = 1$: a Riemann-flat embedding for Milne's universe} 

The choice $h_{00} = 1$ (and $h_{0j} =0$) is usual in cosmology; it corresponds to the so-called synchronous reference system where the time coordinate $t$ is the {\it proper} time at each point. 
In such a system the line element (\ref{solution for k = 2}) becomes
\begin{equation}
dS^2 = dt^2 - \left(B + {C\xi}\right)^2\left[dr^2 + r^2d\Omega^2\right] - \xi^2 dy^2.
\end{equation}
On $\Sigma_{y}$ the effective matter quantities are given by
\begin{equation}
\rho^{(eff)} = - 3 p^{(eff)} = \frac{3C^2}{8 \pi (By  + C t)^2}.
\end{equation}
We note that the equation of state $\rho = - 3p$ appears in different contexts: in Milne's universe; in discussions of premature recollapse problem \cite{Barrow}; in coasting cosmologies \cite{Kolb}; in cosmic strings \cite{Vilenkin4}, \cite{Gott}; limiting configurations \cite{limitConf}; the exterior spacetime for stellar models in $5D$ Kaluza-Klein gravity \cite{Ext.SpacetimeForSrtarsInKK}.

\subsection{Riemann-flat embedding for a $4D$ universe filled with ordinary matter and a cosmological constant}

In order to obtain an equation for the unknown function $\lambda(\xi)$, let us assume that the effective matter can be separated as 

\begin{eqnarray}
8\pi \rho^{eff} &=& 8\pi \rho + \Lambda,\nonumber \\ 
8 \pi p^{eff} &=& 8 \pi p - \Lambda,
\end{eqnarray}
where $\rho$ and $p$ are the density and pressure of the cosmological fluid and $\Lambda$ is the cosmological constant. In addition, as it is usual in cosmology, we assume that the density and pressure satisfy the barotropic equation of state
\begin{equation}
\label{barotropic equation of state}
p = n \rho,
\end{equation}
where $n$ is some constant commonly restricted by $|n| \leq 1$, which follows from the dominant energy condition \cite{Hawking}, \cite{Wald}.
Thus, 
\begin{equation}
\rho = \frac{\rho^{eff} + p^{eff}}{(n + 1)}, \;\;\;\frac{\Lambda(n + 1)}{8\pi} = (n\rho^{eff}- p^{eff}), \;\;\;p = n \rho
\end{equation}
Using (\ref{effective density}) and (\ref{effective pressure}) we find
\begin{equation}
\label{diff. equation for lambda}
2 \lambda_{\xi\xi} + (3n + 2)\lambda_{\xi}^2 - h_{0}\lambda_{\xi}e^{\lambda/2} = 0, \;\;\;\mbox{with} \;\;\;h_{0} \equiv \frac{4\Lambda (n + 1)y^2}{C}.
\end{equation}
where $h_{0}$ is  a constant. Thus, in general 
\begin{equation}
\Lambda =  \frac{h_{0}C}{4(n + 1)y^2}.
\end{equation}
Setting $S(\xi) = e^{\lambda(\xi)/2}$, (\ref{diff. equation for lambda}) becomes
\begin{equation}
\label{equation for S}
2SS_{\xi\xi} + 2(3 n + 1)S_{\xi}^2  - h_{0}S^2 S_{\xi} = 0,
\end{equation}
whose first integral is 

\begin{equation}
\label{First integral for S}
S_{\xi} = a S^2 + \frac{b}{S^{(3n + 1)}}, \;\;\;\mbox{where}\;\;\;a \equiv \frac{2\Lambda y^ 2}{3C},
\end{equation}
i.e., $h_{0} = 6 a(n + 1)$ and $b$ is a constant of integration. Thus, in the case under consideration the spacetime part of the $5D$ metric (\ref{solution for k = 2}) is given by

\begin{equation}
ds^2 = \frac{2}{C }\left(a S^2 + \frac{b}{S^{(1 + 3 n)}}\right)dt^2 - S^2\left[dr^2 + r^2\left(d\theta^2 + \sin^2\theta d\phi^2\right)\right].
\end{equation}
The corresponding matter quantities are
\begin{equation}
\label{matter quantities for S}
8\pi\rho = \frac{3C b}{2y^2 S^{3(n +1)}}, \;\;\;p = n \rho, \;\;\;\Lambda = \frac{3C a}{2y^2}.
\end{equation}
Let us consider the deceleration parameter $q$, which is defined as
\begin{equation}
\label{q}
q = - \frac{S_{\tau \tau}S}{S_{\tau}^2},
\end{equation}
where $S_{\tau} = (1/y)S_{\xi}e^{- \nu/2}$ represents the derivative of the scale factor with respect to the universal  time $\tau$, which is related to the coordinate time $t$ by the expression $d\tau = e^{\nu/2}dt$. A simple calculation gives,
\begin{equation}
\label{specific form for q}
q = \frac{3n + 1}{2} - \frac{3 a\;(n + 1)}{2[a + b/S^{3(n + 1)}]}.
\end{equation}
In order to simplify this expression let us  calculate the density parameters $\Omega_{m} = 8\pi \rho/3H^2$ and $\Omega_{\Lambda} =\Lambda/3H^2$, where $H \equiv S_{\tau}/S$ is the Hubble parameter. We obtain
\begin{equation}
\label{cosmological parameters}
H^2 = \left(\frac{Ca}{2y^2}\right)\left[1 + \frac{(b/a)}{S^{3(n + 1)}}\right], \;\;\;\;\Omega_{m} =  \frac{(b/a)}{S^{3(n + 1)} + (b/a)}, \;\;\;\;\Omega_{\Lambda} = \frac{S^{3(n + 1)}}{S^{3(n + 1)} + (b/a)}.
\end{equation}
Consequently,
\begin{equation}
S^{3(n + 1)} = \left(\frac{\Omega_{\Lambda}}{\Omega_{m}}\right)\left(\frac{b}{a}\right).
\end{equation}
Substituting this expression into (\ref{specific form for q}), and using that $\Omega_{m} + \Omega_{\Lambda} = 1$, we get
\begin{equation}
q = \frac{3}{2}\Omega_{m}(n + 1) - 1.
\end{equation}
We note that $q$ changes sign at $\Omega_{m} = \Omega^{(crit)}_{m} = 2\Omega_{\Lambda}/(3n + 1)$: for $\Omega_{m} < \Omega^{(crit)}_{m}$
the expansion is slowing down $(q > 0)$, for $\Omega_{m} > \Omega^{(crit)}_{m}$ the expansion is speeding up $(q < 0)$. 
Setting $n = 0$, in concordance  with the fact that our present universe is matter-dominated $(p = 0)$, and $\Omega_{m} = {\Omega_{m}}_{|today} \approx 0.3$ we obtain the approximate value of deceleration parameter today, viz.,
\begin{equation}
\label{present value of q}
q_{|today} \approx - 0.55,
\end{equation}
which is consistent with observations \cite{Bernandis}, \cite{Spergel2}.

\subsubsection{Analysis of the evolution equation (\ref{First integral for S})}

Let us notice that, in general,  (\ref{First integral for S}) cannot be integrated in terms of elementary functions. However the 	approximate behavior of the solution is as follows. 
In the very early universe, when $S \approx 0$, the second term in the r.h.s. of (\ref{First integral for S})
dominates over the first one. Thus, we find $S \sim \xi^{1/(3n + 2)}$ which implies  that the actual behavior of the scale factor depends on the equation of state.

\paragraph{False vacuum: n = - 1.} Following Zeldovich \cite{Zeldovich}, one may imagine that after the spontaneous  birth (where the spacetime separates  from the extra dimension)  our universe enters a de Sitter phase of expansion. In this phase  $n = -1$ and $S \sim 1/t$ on every hypersurface $\Sigma_{y}$, which  means that $S = 0$ at coordinate time $t = \infty$. With the transformation of coordinates $t \rightarrow e^{- \omega \tau}$, where $\omega$ is a constant, we obtain $S \sim e^{\omega \tau}$. Thus, $S = 0$ at $\tau = - \infty$ which corresponds to $t = \infty$.  Thus, during the inflationary period  the expansion of the universe is described by 
\begin{equation}
\label{inflation}
ds^2 = d\tau^2 - A e^{2\sqrt{\Lambda_{(infl)}/3}\tau}\left[dr^2 + r^2d\Omega^2\right],
\end{equation}
where $\Lambda_{(infl)} = 3\omega^2$ is the ``effective" cosmological constant in this epoch. According to  (\ref{matter quantities for S})  it is related to the present cosmological constant by 
\begin{equation}
\label{Lambda for inflation}
8\pi \rho^{(eff)} = \Lambda_{(infl)} = \Lambda \left(1 + \frac{b}{a}\right).
\end{equation}

However, it should be noted that the de Sitter  solution is unstable under small perturbations. Therefore, it is impossible to extrapolate it to $\tau = - \infty$ \cite{Zeldovich}. What this means  is that (\ref{inflation}) does not describe  $S = 0$, which in singular on $\Sigma_{y}$ (but non-singular in $5D$),  corresponding to the moment of spontaneous  birth of our universe\footnote{Models for  inflation estimate that the de Sitter phase of exponential expansions begins at $\tau = 10^{- 42}$ s \cite{LindeBook}. }.

\paragraph{FRW evolution.}

The switch from the de Sitter exponential expansion to the radiation dominated FRW universe is a jump of pressure from $p = - \rho$ to $p = \rho/3$. In this case, $S \sim t^{1/3}$ on every $\Sigma_{y}$.  Now changing $t \rightarrow \tau^{3/2}$ the scale factor becomes $S \sim \tau^{1/2}$, which is the usual expression for a radiation dominated era. For any $n \neq - 1$ the transformation  $t \rightarrow \tau^{2(3n + 2)/[3(1 + n)]}$ allows us to  recover the familiar  spatially-flat FRW models, viz., 
\begin{equation}
\label{FRW models}
ds^2 = d{{\tau}}^2 - B \tau^{4/[3(1 + n)]}\left[dr^2 + r^2d\Omega^2\right].
\end{equation}
Let us call $\tau_{0}$ the moment of transition from the de Sitter phase to the relativistic plasma $(n = 1/3)$ FRW solution. A jump in the pressure indicates that the second derivative of the scale factor is discontinuous, which in turn  implies  a jump in the deceleration parameter $q$;  in the present case  from $q = - 1$ to $q = 1$. However, in order to guarantee the continuity of energy density and the Hubble term, 
the scale factor and its first derivative must be continuous during the jump\footnote{This is equivalent to requiring continuity of the first and second fundamental forms across a $3D$ surface $\tau = \tau_{0} = $ constant \cite{JPdeLDynamicalFoliationII}. }. This allows us to relate $\tau_{0}$ and $\Lambda_{(infl)}$, namely,

\begin{equation}
\label{jump time}
\tau_{0} = \frac{1}{2}\sqrt{\frac{3}{\Lambda_{(infl)}}}.
\end{equation}  
Our toy model does not provide any mechanism for calculating the time of transition, neither is the object of the present work, however various inflationary models suggest that inflation ends at $\tau_{0} = 10^{- 32 \pm 6}$ s \cite{LindeBook}. Using this number as a reference we obtain

\begin{equation}
\label{evaluating Lambda during inflation}
\Lambda_{(infl)} \sim 10^{48 \pm 12} cm^{- 2}, 
\end{equation}
which is huge compared to the present value of the cosmological constant $\Lambda \sim 10^{-51}$ cm$^{- 2}$, measured in the current $\Lambda$CDM model of cosmology \cite{Bernandis}-\cite{Spergel2}. Thus,    
\begin{equation}
\label{very large ratio}
\Lambda_{(infl)} \approx 10^{99 \pm 12} \Lambda.
\end{equation}
 
\paragraph{Accelerated expansion.}  As the universe expands the deceleration parameter $q$, which is given by (\ref{specific form for q}), changes sign again at 
\begin{equation}
\label{S for q = 0}
S_{(q = 0)} = \left[\frac{3n + 1}{2}\left(\frac{\Lambda_{(infl)}}{\Lambda} - 1\right)\right]^{1/3(n + 1)},
\end{equation}
where we have used (\ref{Lambda for inflation}) to express the ratio $(b/a)$. This expression uncovers two interesting physical concepts: (i) if $\Lambda_{(inf)} = \Lambda$, then  the universe would have never changed from decelerated to accelerated expansion, and (ii) the very large ratio $\Lambda_{(infl)/}/\Lambda$, evaluated in (\ref{very large ratio}), explains the very large radius of the universe.

The present value of $q$, as calculated in (\ref{present value of q}),  is approximately  $- 0.55$ which means that our universe is in a phase of accelerated expansion.  At late times, for large values of $S$, $ (S > S_{(q = 0)})$, the first term in (\ref{First integral for S}) starts dominating over the second one. Asymptotically, for $S \gg S_{(q = 0)}$, in terms of universal time $\tau$ our world enters a new era of exponential  expansion with $q = - 1$ 
and 
\begin{equation}
\label{accelerated expansion}
ds^2 = d\tau^2 - e^{2\sqrt{\Lambda/3}\tau}\left[dr^2 + r^2d\Omega^2\right].
\end{equation}
In summary, the toy cosmological models considered in this section allow us to conclude that that the Riemann-flat embeddings given by (\ref{self-similar first solution}) are rich enough as to 	accommodate the essential features of the evolution of our universe.

\section{Concluding remarks}

In this paper we have found the most general self-similar, Ricci-flat, homogeneous and isotropic cosmologies in $5D$. Self-similarity requires that all the dimensionless quantities in the theory be functions of a single variable $\xi$, which in the case studied here is some combination  of $t$ and $y$ (\ref{The most general self-similar variable}). 
As a consequence of the  symmetry, and the field equations $R_{AB} = 0$,  we have found that 
(i) the three-dimensional spatial surfaces defined by $t =$ constant and $y = $ constant have to be flat (\ref{most general self-similar line element 1});  (ii)  the self-similar variable can be taken either as $\xi = t/y$ (\ref{self-similar first solution}), which is the more general case, or as $\xi = e^{\alpha t}/e^{\alpha y}$ (\ref{self-similar first solution 2}); (iii) the extra dimension must be spacelike (\ref{Phi for k neq 1}). Then, we obtained the general solutions of the field equations in terms of one arbitrary function of $\xi$ (\ref{self-similar first solution})-(\ref{self-similar first solution 2}). 

These  solutions are Riemann-flat in $5D$ but curved in $4D$ (\ref{Riemann tensor in 4D}). This latter property, together with the property (i) mentioned above, allow us to interpret these cosmologies as Riemann-flat embeddings for spatially-flat  FRW cosmologies in $4D$. In this interpretation our universe arises as a spontaneous separation from an empty $5D$ Minkowski space, as envisioned by Zeldovich \cite{Zeldovich}.

We have analyzed in some detail the $4D$ interpretation  of (\ref{self-similar first solution}) (for $l = 2$), which admits the simply homothethic Killing vector $\xi^{A} = (t, \; r, \; 0,\; 0, \; y)$. We have seen that the $4D$ projection of this vector does not constitute a $4D$ homothetic Killing vector for the  metric induced on $\Sigma_{y}$ (\ref{lost of symmetry in 4D}).  For the physics  in $4D$ this opens a wide range of non-homothetic, but still self-similar,  possibilities for the evolution of our universe (\ref{self-similar metric on Sigma})-(\ref{Lie derivative of the metric}), among them it allows the introduction of a cosmological constant and a de Sitter phase of exponential expansion. 

The question is how to establish the arbitrary function of $\xi$ in (\ref{self-similar first solution})-(\ref{self-similar first solution 2}). There are many plausible ways for selecting it. As an  example one can choose some ``geometrical" criteria, e.g.,  choose a particular  coordinate frame or to assume some specific symmetry, as we do in sections $4.1$, $4.2$ and the Appendix, respectively.  A distinct approach, which is probably more satisfactory from a ``physical" point of view,   is to select the arbitrary function  by imposing conditions on the effective picture in $4D$; this is what we do in section $4.3$. Certainly, the existence of an arbitrary function allows us to accommodate a number of physical models.

The model discussed in $4.3$ roughly presents  the essential features of the evolution of our universe. Specifically, it shows inflation from an original birth, not described by the de Sitter solution (\ref{inflation}), followed first by a FRW  phase and then by an accelerated expansion. The effective cosmological constants during inflation $\Lambda_{(infl)}$ and after inflation $\Lambda$ are not equal, otherwise the universe cannot enter a phase of accelerated expansion (\ref{S for q = 0}).

  In summary, in this work we have studied a family of self-similar cosmological metrics in $5D$. As far as the author knows, this is the first work where self-similarity is used for finding exact solutions to the field equations in $5D$. The solutions may be used as $5D$ embeddings for the spatially-flat FRW cosmological models of ordinary general relativity in $4D$.

We have not investigated fully the possible physical interpretations of (\ref{self-similar first solution}), we have just considered a particular  case. Neither have we investigated the solution (\ref{self-similar first solution 2}), which seems to be totally different from  (\ref{self-similar first solution}).
An immediate extension of this work is the study of self-similar cosmologies with extra dimensions {\it without} the assumption of spatial isotropy and/or homogeneity.   Also, it is interesting to find the transformation of coordinates from 
(\ref{self-similar first solution})-(\ref{self-similar first solution 2}) to ${\cal{M}}_{5}$ given by (\ref{Minkowski space in 5D}). This is a non-trivial task, but the existence of such a transformation is guaranteed by the fact that the Riemann tensor in $5D$ vanishes. With the appropriate transformation of coordinates at hand one can extend and generalize  previous investigations \cite{SanjeevWesson} and study the birth of the universe in more detail.

\renewcommand{\theequation}{A-\arabic{equation}}
  \setcounter{equation}{0}  
  \section*{Appendix: Homothetic symmetry on $\Sigma_{y}$}  

Our aim here is (i) to show that the requirement of homothetic symmetry on $\Sigma_{y}$ singles out one specific metric in $4D$, namely, the spacetime part of the $5D$ metric (\ref{standard cosmological solution in 5D in self-similar form}), and that (ii) the homothetic vector is not parallel to $\xi^{\mu}_{p}$ (\ref{lost of symmetry in 4D}). 

In order to do this, let us express  the induced metric on $\Sigma_{y}$, which is given by (\ref{4D metric}), in terms of $S = e^{\lambda/2}$,  
\begin{equation}
\label{self-similar metric on Sigma}
ds^2 = h_{\mu\nu}dx^{\mu}dx^{\nu} = \frac{2S_{\xi}}{C}dt^2 - S^2(\xi)\left[dr^2 + r^2 d\Omega^2\right]. 
\end{equation}
The Lie derivative of this metric along the $4D$ vector
\begin{equation}
\label{homothetic Killing vector}
\zeta^{\mu} = (At, \;Br, \;0, \;0),
\end{equation}
where $A$ and $B$ are some constants, is given by
\begin{equation}
\label{Lie derivative of the metric}
 {\cal{L}}_{\zeta} h_{0 0}  = 2h_{00}\;A\left(1 + \frac{\xi S_{\xi \xi}}{2 S_{\xi}}\right), \;\;\;                                      {\cal{L}}_{\zeta} h_{i j} = 2 h_{i j}\left[A\frac{\xi S_{\xi}}{S} + B\right] 
\end{equation}
Thus, the requirement 
\begin{equation}
\label{homothetic requirement}
{\cal{L}}_{\zeta}h_{\alpha \beta } = 2 h_{\alpha\beta},
\end{equation}
generates two independent equations, viz., 
\begin{eqnarray}
\frac{S_{\xi\xi}}{S_{\xi}} = \frac{2(1 - A)}{A\; \xi}, \;\;\;\mbox{and}\;\;\;\frac{S_{\xi}}{S} = \frac{(1 - B)}{A\;\xi} .
\end{eqnarray}
From the first equation we get $ S = C_{1}\xi^{(2 - A)/A} + C_{2}$, while from the second one we obtain $S = C_{3}\xi^{(1 - B)/A}$, where $C_{1}$, $C_{2}$ and $C_{3}$ are constants of integration. Compatibility of these expressions  demand $(A - B) = 1$, and  $C_{2}$ = 0. Consequently, $S \sim \xi^{(2 - A)/A}$ and $e^{\nu} \sim \xi^{2(1 - A)/A}$. 

Since $A$ is an arbitrary parameter, without loss of generality we can set
\begin{equation}
A = \frac{2\alpha - 1}{\alpha}.
\end{equation}
With this selection the metric in $4D$  becomes
\begin{equation}
\label{homothetic solution on Sigma}
ds^2 = A \xi^{2(1 - \alpha)/(2\alpha - 1)}dt^2 - B \xi^{2/(2\alpha - 1)}\left[dr^2 + r^2d\Omega^2\right],
\end{equation}
which is identical to the spacetime part of (\ref{standard cosmological solution in 5D in self-similar form}) and admits  the  homothetic Killing vector  
\begin{equation}
\label{homothetic vector on Sigma}
\zeta^{\mu} = \left(\frac{2\alpha - 1}{\alpha}t, \; \frac{\alpha - 1}{\alpha}r, \; 0, \; 0\right).
\end{equation}
Clearly, this vector  is not parallel to $\xi^{\mu}_{p} = (t, \; r, \; 0, \; 0)$. We emphasize that the $5D$ metric (\ref{5D homothetic vector in self-similar coordinates}) is homothetic along the $5D$ vector $\xi^{A} = (t, \;r, \; 0, \; 0, y)$, but its spacetime part is homothetic along (\ref{homothetic vector on Sigma}) and not along $\xi^{\mu}_{p} = (t, \; r, \; 0, \; 0)$.

\end{document}